# Unraveling the pH-Dependent Oxygen Reduction Performance on Single-Atom Catalysts: From Single- to Dual-Sabatier Optima

Di Zhang[1,3], Zhuyu Wang[2], Fangzhou Liu[2], Peiyun Yi[3], Linfa Peng[3], Yuan Chen[2], Li Wei[2,*], and Hao Li[1,*]

[1] Advanced Institute for Materials Research (WPI-AIMR), Tohoku University, Sendai, 980-8577, Japan,

[2] School of Chemical and Biomolecule Engineering, The University of Sydney, Sydney, New South Wales, 2006, Australia

[3] State Key Laboratory of Mechanical System and Vibration, Shanghai Jiao Tong University, Shanghai, 200240, China

* Corresponding Authors:

Email: l.wei@sydney.edu.au (L. W.)

Email: li.hao.b8@tohoku.ac.jp (H. L.)

ABSTRACT

Metal-nitrogen-carbon (M-N-C) single-atom catalysts (SACs) have emerged as a potential substitute for the costly platinum-group catalysts in oxygen reduction reaction (ORR). However, several critical aspects of M-N-C SACs in ORR remain poorly understood, including their pH-dependent activity, selectivity for 2- or 4-electron transfer pathways, and the identification of the rate-determining steps. Herein, by analyzing >100 M-N-C structures and >2000 sets of energetics, we unveil a pH-dependent evolution in ORR activity volcanos—from a single-peak in alkaline media to a double-peak in acids. We found that this pH-dependent behavior in M-N-C catalysts fundamentally stems from their moderate dipole moments and polarizability for O* and HOO* adsorbates, as well as unique scaling relations among ORR adsorbates. To validate our theoretical discovery, we synthesized a series of molecular M-N-C catalysts, each characterized by well-defined atomic coordination environments. Impressively, the experiments matched our theoretical predictions on kinetic current, Tafel slope, and turnover frequency in both acidic and alkaline environments. These new insights also refine the famous *Sabatier* principle by emphasizing the need to avoid an "acid trap" while designing M-N-C catalysts for ORR or any other pH-dependent electrochemical applications.

1. INTRODUCTION

Metal-nitrogen-carbon (M-N-C) single-atom catalysts (SACs) have recently garnered significant attention.[1-9] They are highly promising due to their encouraging catalytic activity for the oxygen reduction reaction (ORR) and much lower costs compared with Pt-group metal (PGM) catalysts that are typically used to overcome the sluggish ORR kinetics in fuel cells and metal-air batteries. In recent years, there has been remarkable progress in the design,[2-4] synthesis,[5] characterization,[6,7] catalytic activity, and stability evaluation[1,8,9] of M-N-C SACs. Despite these advancements, the scientific community in this field has yet to reach a unified understanding of the pH dependence, the preference of the 4-/2-electron (4e⁻/2e⁻) selectivity, and rate-determining steps of different M-N-C SACs, as highlighted in recent works.[10-15]

To showcase the versatility of M-N-C catalysts, **Figure 1** summarizes the reported experimental ORR activities and 4e⁻/2e⁻ selectivity of >100 M-N-C catalysts in both acidic and alkaline media. These catalysts feature different center metals, including Mn, Fe, Co, Ni, and Cu, as well as diverse surrounding functional groups[16-22] (for more details, see **Table S1-S2**). These M-N-C catalysts include those prepared by conventional high-temperature pyrolysis of hybrid precursors constituted from carbon matrices (*e.g.*, graphene or carbon nanotube, CNT) and metal-containing complexes.[23] We also incorporated molecular M-N-C catalysts, such as metal-organic frameworks (MOF)-derived catalysts (*e.g.*, porphyrin (POR), and phthalocyanine (Pc)) and covalent organic framework (COF)-based catalysts, which have attracted broad interests in recent years due to the promoting dispersion effect for metal sites and the presence of more identical neighboring coordinated atoms around metal centers.[21, 22] In light of **Figure 1**, three fundamental questions emerge pertaining to the pH-dependent behavior of M-N-C catalysts:

- Why do M-N-C catalysts exhibit varying pH dependence? For instance, **Figure 1** indicates that certain Fe-based M-N-C catalysts appear to have less pH dependence when compared to catalysts centered around other metals. Even when the catalysts are both Fe-centered, how can we explain the better ORR performance observed on Fe-pyrrole-$N_4$ compared to Fe-pyrridine-$N_4$ in acidic conditions? [1, 24]

- How does pH affect the ORR selectivity of M-N-C catalysts? As depicted in **Figure 1**, what factors contribute to the generally higher selectivity for $H_2O_2$ observed in Co-centered M-N-C catalysts,[22] and why is this selectivity further intensified in acidic conditions? Why do pyrolysis M-N-C catalysts generally exhibit a lower $H_2O_2$ Faradaic efficiency (FE) compared to molecular M-N-C catalysts?

- What is the origin of the versatility of M-N-C SACs? To elaborate, some M-N-C catalysts exhibit insignificant pH dependence and 4e$^-$-ORR selectivity comparable to PGMs in both alkaline and acidic media.[6] In the meantime, others manifest a pronounced pH sensitivity and higher FE towards $H_2O_2$, like many of transition metal X-ides (TMXs, X=C, N, O, etc.).[25]

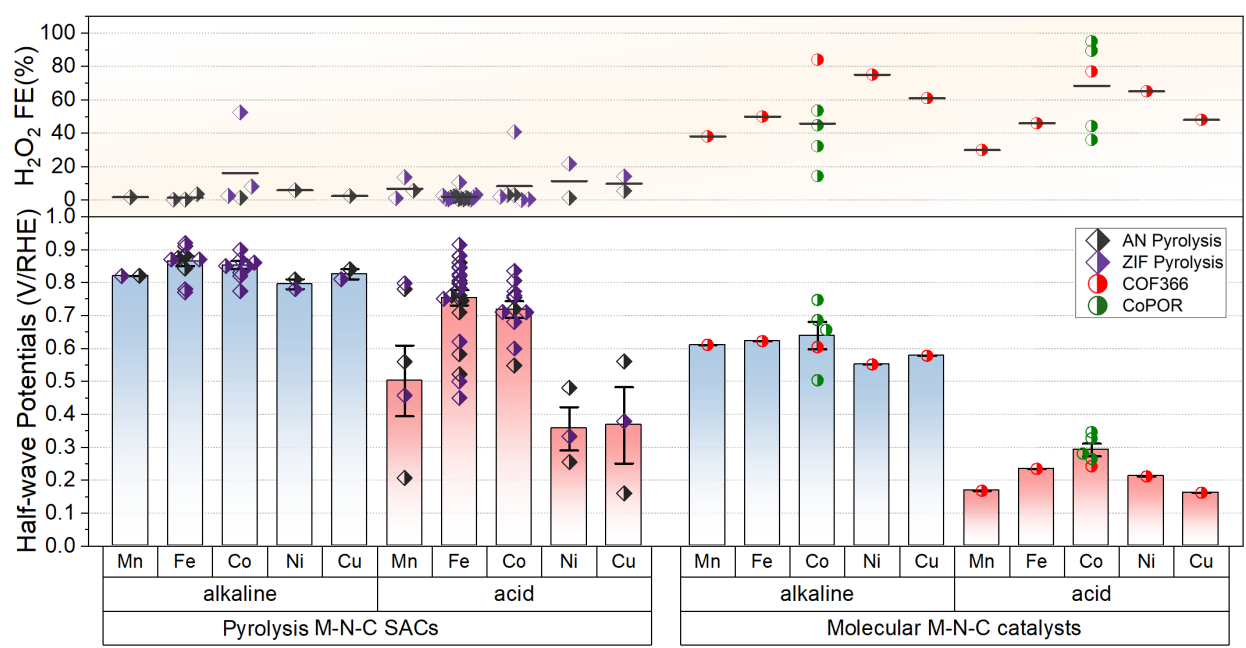

**Figure 1. ORR activity analysis of >100 reported M-N-C catalysts.** An overview of typical M-N-C catalysts for ORR with detailed data sources available in **Table S1-S2**. Data points were extracted from the original references based on the following conditions: the half-wave potential is the potential at which the current reaches half the value of the diffusion limiting current, and the Faradic efficiency (FE) is calculated at 0.6 V *vs.* reversible hydrogen electrode (RHE). For COF and POR-based catalysts, the potential at a current density of 0.5 mA cm$^{-2}$ is extracted as the half-wave potential because no diffusion-limiting current can be observed. The FEs of COF and POR-

based catalysts are calculated at 0.30 and 0.48 V/RHE, respectively. More details related to previous experimental data can be found in the **Supplementary Information**.

To date, one of the most widely employed approaches to model electrocatalysis is the computational hydrogen electrode (CHE) method[26] based on density functional theory (DFT) calculations. Although conventional thermodynamic models can capture the trends for the 2e$^-$ and 4e$^-$ ORR,[27] accurately predicting experimental observations, such as current density ($j$), half-wave potential ($E_{1/2}$), and Tafel slope, remains difficult without microkinetic modeling. Moreover, modeling the pH-dependent activity and selectivity of M-N-C catalysts presents a tougher challenge due to the absence of pH/potential-dependent free energies.[2] Recently, the origin of the high ORR activity of Fe-based M-N-C catalysts was considered to be the Fe-pyrrole-N species through DFT calculations.[28] Another theoretical study also analyzed the ORR activity of Fe-N-C catalysts by using various descriptors, such as the Fe-O bond length, the $d$-band center gap of spin states, and the magnetic moment of Fe-site and $O_2$*.[29] Regarding the remarkable $H_2O_2$ selectivity of Co-base M-N-C catalysts, a recent model based on constant-potential explicit-solvation *ab initio* molecular dynamics (AIMD) was developed to explain the origin of selectivity, specifically the proton adsorption to the former or later O in HOO*.[11] Despite the insights provided by these models, there is still a lack of an accurate prediction of pH-dependent activity and selectivity. It is worth noting that a recent study[15] explored the electric field effects on the adsorption strengths of ORR intermediates, providing insights into why the pyrrole Fe-N-C in acidic media exhibits a higher activity than pyridinic Fe-N-C catalysts; however, the predicted pH-dependent activities differ from experimental observations regarding the onset potential under acidic and alkaline media. The urgent needs for a unified model that could accurately depict the pH-dependent ORR landscape of M-N-C catalysts remain as pressing as ever.

## 2. METHODS

**Computational methods.** For the binding energies of ORR adsorbates, DFT calculations were performed using the generalized gradient approximation method with the revised Perdew−Burke−Ernzerhof functional to describe electronic exchange and correlations.[30, 31] A projector augmented-wave method was used to describe the core electrons.[32] Valence electrons were described by expanding the Kohn–Sham wavefunctions in a plane-wave basis set,[33] with a cutoff of at least 400 eV. We used the Quantum Espresso code[34] to calculate the electric field effects. Electric fields were applied using a saw-tooth potential that corresponds to fields ranging from −0.4 to 1.2 V/Å. At each applied field, adsorbates were allowed to relax with a force convergence threshold of 0.05 eV/Å. We used the lowest energy conformation to predict the adsorbate energy under that field. The complete computational and modelling details can be found in the **Supplementary Information**.

**Synthesis of M-COF366/CNT and MPc/CNT catalysts.** Materials and purification of as-received CNTs can be found in **Supplementary Information**. The diameter of the CNTs is about 10~20 nm. The effect of CNT curvature on the catalytic performance can be ignored when the curvature falls within a range of 10-20 nm and the diffusion of molecular catalysts into the inner CNT is kinetically unfavored (see **Figure S1** and **Figure S2** for more details). The COF-366 decorated MWCNT catalysts was prepared by a solvothermal method. About 0.02 mmol TAPP and 0.04 mmol TPD were added to a mixture of 1 mL of absolute ethanol, 1 mL of mesitylene, and 0.1 mL of 6 M acetic acid in a Pyrex tube. Purified CNT was added at a mass ratio of 2:1 to the COF-366 precursors. The mixture was then suspended and mixed by sonication for 30 min. The tube was then subjected to three-round freeze-thaw cycles using liquid $N_2$, preheated to 65 °C for 4 hours under Ar protection before further hydrothermally treated at 120 °C for 72 hours. The solid products were then recovered by filtration and washed with EtOH and DMAC repeatedly before being dried in a vacuum. The product is denoted as COF366/CNT and serves as the substrate to prepare M-COF366/CNT catalysts. We repeated these synthesis processes to accumulate enough material for the following metalation step. Metalation of the COF366/CNT was then performed using various metal acetate salts as metal precursors. About 45 mg of the as-prepared COF366/CNT substrate and 0.27 mmol of the metal salts were suspended in 5 mL MeOH, followed by adding 20 mL $CHCl_3$ and 15 mL DMF. The mixture was bath sonicated for 60 min and further stirred at 80 °C under Ar protection for 24 hours. After cooling to room temperature, the solid product was recovered by filtration and washed with DI water before being dried in a vacuum. During this preparation process, the COF366 was firstly synthesized to coat the CNT substrate, forming a core-shell structure. Afterwards, the metal was loaded to afford the M-COF366/CNT catalyst. This method could minimize the metal infiltration to inner porphyrin moieties buried inside and avoid the stacking of metal active sites. The resulting catalysts were denoted as M-COF366/CNT, where M is Fe, Co, Ni, and Cu.

The MPc/CNT catalysts were synthesized by loading the various MPc (M=Fe, Co, Ni, and Cu) and FePc molecules with different R-groups (–R = –H, –$NH_2$, –$NO_2$, –F, and –*tert*-butyl) on the purified CNT substrate *via* van der Waals interactions following our reported method.[22] Briefly, about 3 mg of the MPc compounds and 20 mg of purified MWCNT were dissolved in 20 mL DMF by ultrasonication for 30 min. Afterward, the mixture solution was stirred under an Ar environment for another 24 hours before the solid products were filtered and washed with DMF (3×15 mL) and ethanol (10 mL) before they were dried under vacuum at 80 °C overnight to obtain the MPc/CNT catalysts.

**Characterization and Electrochemical Methods.** Thermogravimetric analysis (TGA) was performed on a TA Instrument TGA 5500 thermo analyzer under airflow (20 sccm). Liquid $N_2$ physisorption isotherms were collected on an Anton Paar Autosorb iQ analyzer. Metal residue in the purified CNT and metal loading in various catalysts were determined by inductively coupled plasma atomic emission spectroscopy (ICP-AES) on a Perkin Elmer Avio 500 spectrometer. The sample was acid digested in 6 M $HNO_3$ before dilution and measurement. X-ray photoelectron spectra (XPS) were obtained on a Thermo Scientific K-Alpha+ spectrometer with an Al-Kα source (1486.3 eV). Samples were loaded on a gold substrate. Survey spectra are obtained at a 1 eV step in CAE mode, with a pass energy of 200.0 eV. The high-resolution spectra were collected at a step of 0.1 eV. Transmission electron microscope (TEM) images and energy-dispersive X-ray (EDX) elemental analysis are collected on an FEI Themis-Z microscope, under either bright-field high-resolution (BF-HRTEM) or high-angle angular dark-field scanning mode (HAADF-STEM) modes. Synchrotron illuminated X-ray absorption spectra (XAS) were collected at the XAS Beamline at the Australian Synchrotron (ANSTO). Electron beam was collected from a set of liquid nitrogen cooled Si(111) monochromator and associated Si-coated collimating and Rh-coated focusing mirrors. The beam size was about $1 \times 1$ $mm^2$. Data was collected under the transmission mode, and the energy was calibrated using corresponding metal foils. Afterwards, the spectra were processed and fitted in the Demeter package using FEFF 9.0 code.

The electrochemical performance of different catalysts was collected using a CHI760 electrochemical workstation and a Pine MSR rotator in a three-electrode configuration at 25 °C. A rotary ring-disk electrode (RRDE, E6R1, Pine Research, with a calibrated collection efficiency $N$=0.249) equipped with a glassy carbon disk (GCD, 5 mm diameter) and a Pt ring (OD = 7.50 mm, ID = 6.50 mm). The electrode was polished using $Al_2O_3$ powder before each measurement.

A pre-calibrated Ag/AgCl (3M KCl filling) reference electrode and a Hg/HgO (0.1 M KOH filling) reference electrode were used for the acidic 0.1 M $HClO_4$ and the alkaline 0.1 M KOH electrolytes, respectively. A graphite rod electrode (AFCTR3B, Pine Research) was used as the counter electrode. All reported potentials were calibrated to a reversible hydrogen electrode ($V_{RHE}$). The calculation method for the kinetic current densities and TOFs can be found in **Supplementary Methods**.

3. RESULTS AND DISCUSSION

In this study, we take the crucial first step towards understanding M-N-C SACs by DFT calculations that uncover their unique scaling relations among the adsorbate bonding strengths. 103 M-N-C catalysts with varied center metal atoms, number of coordinating N atoms, and surrounding functional groups were analyzed, including M-pyridine-$N_{4-1}$ and M-pyrrole-$N_{4-1}$ (**Figure 2a**), M-COF-366 (**Figure 2b**), M-phthalocyanine-R (M-Pc-R, **Figure 2c**), and Co-porphyrin-R (Co-POR-R, **Figure 2d**). **Figure 2e-g** presents the data collected for the respective scaling relations between the adsorption energies of O*, HOO*, and $O_2$* against HO*. One key finding is that the scaling relation between O* and HO* in M-N-C catalysts falls between that of transition metals (TMs) and TMXs computed based on the same level of accuracy. Interestingly, M-N-C catalysts with unsaturated nitrogen atoms (*i.e.*, 1, 2, or 3 N atoms bonded to center metals) exhibit a lower intercept (0.7) compared to $N_4$-coordinated M-N-C catalysts (1.26) with the same slope of 1.5 (**Figure 2e**). A lower slope indicates stronger adsorption of ORR adsorbates on M-N-C catalysts, which is attributed to the unsaturated N coordination of the metal atom. A higher intercept indicates a weaker bonding of O* at a given HO* bonding strength, which in turn results in a larger barrier for O-O bond breaking.[25]

The second scaling relation (HO* vs. HOO*) observed in **Figure 2f** appears to be universal, with an intercept nearly identical to those of TMs and TMXs.[25] A noteworthy discovery is that $E_{O_2*}$ and $E_{HO*}$ display a segmented relationship in **Figure 2g**: in the M-N-C catalysts with $E_{HO*}$ values surpassing 1.5 eV, the binding energy of $O_2*$ remains nearly constant at 4.95 eV due to the physical adsorption state, as depicted in the inset image of **Figure 2g**. The scaling relation between $E_{O_2*}$ and $E_{HO*}$ plays a crucial role in HOO* formation during ORR, providing an opportunity to identify the turning point from the 4e⁻ to 2e⁻ ORR selectivity. All of these are borne out in the universal microkinetic model of ORR, which will be discussed in **Figure 4**. Because all reaction energetics have good scaling relations with $E_{HO*}$, a single descriptor is sufficient for microkinetic modeling.[35, 36]

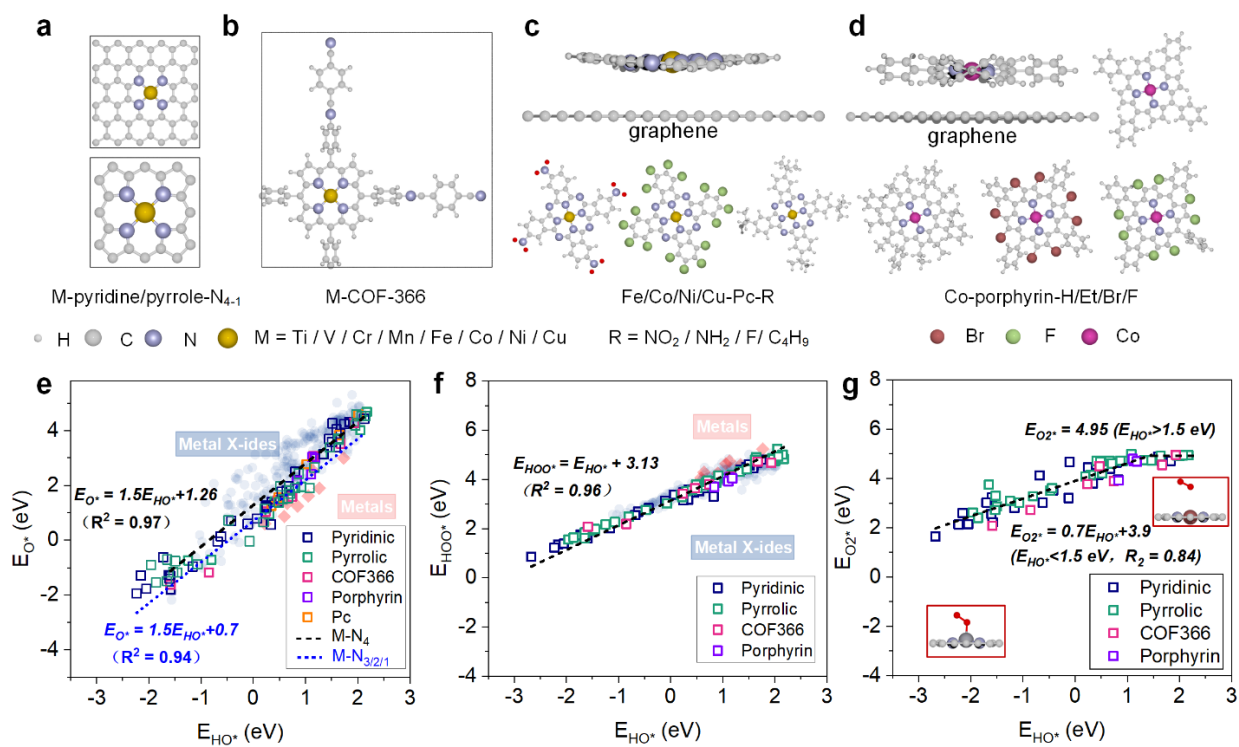

**Figure 2. Linear scaling relations found in M-N-C catalysts with different atomistic environments.** (**a**) One to four pyridine nitrogen atoms (M-pyridine-$N_{4-1}$) and pyrrole nitrogen atoms (M-pyrrole-$N_{4-1}$). (**b**) Conjugated porphyrin-based M-COF-366 SACs. (**c**) M-

phthalocyanine-R. (**d**) Co-porphyrins-R, where R indicates different surrounding functional groups. (**e**) $E_{HO*}$ vs. $E_{O*}$ scaling relation on the above M-N-C catalysts. (**f**) The universal scaling relation of $E_{HO*}$ vs. $E_{HOO*}$. (**g**) $E_{HO*}$ vs. $E_{O_2*}$ scaling relation on the metal site of M-N-C catalysts. Insets show the identified chemical and physical adsorption states.

To develop a comprehensive model, we need to examine all relevant factors that can have a significant impact on the ORR activity of M-N-C SACs, including the potential of zero charges (PZCs), and the electric field and solvation effects. Recent studies have demonstrated that electric field effects can account for the pH dependence of weak-binding ORR catalysts.[35, 37] Although it may be difficult to accurately measure the magnitude of the electric field under different electrode potentials and pH, a parallel-plate capacitor model can be used to describe the relationship between the electric field and electrode potential (*vs.* standard hydrogen electrode, SHE).[35] The field interacts with intermediates and transition states with a substantial dipole moment ($\mu$) and/or polarizability ($\alpha$). **Figure 3a-c** provides an example of how ORR adsorbates on TMs,[35] M-N-C catalysts, and TMXs[25] respond to varying electric fields. In accordance with the observed scaling relations between $E_{O*}$ and $E_{HO*}$, M-N-C catalysts demonstrate a moderate sensitivity (especially for the adsorbed O*) to the perturbations in electric field, as depicted by the arrow in **Figure 3**, when compared to this characteristic of TMX[25] and TM surfaces.[35] In **Figure S2-S3**, we delve into the field response of 24 distinct M-N-C catalysts, each differing in metal centers and N coordination numbers. We found the majority of M-N-C catalysts have a $\mu(O*)$ ranging between 0.3 and 0.6 eÅ, which notably situated between the $\mu(O*)$ values observed for TMs and TMXs. Subsequently, we'll highlight how this moderate response, combined with special scaling relations, plays an essential role in the distinct pH dependence and selectivity seen in M-N-C catalysts.

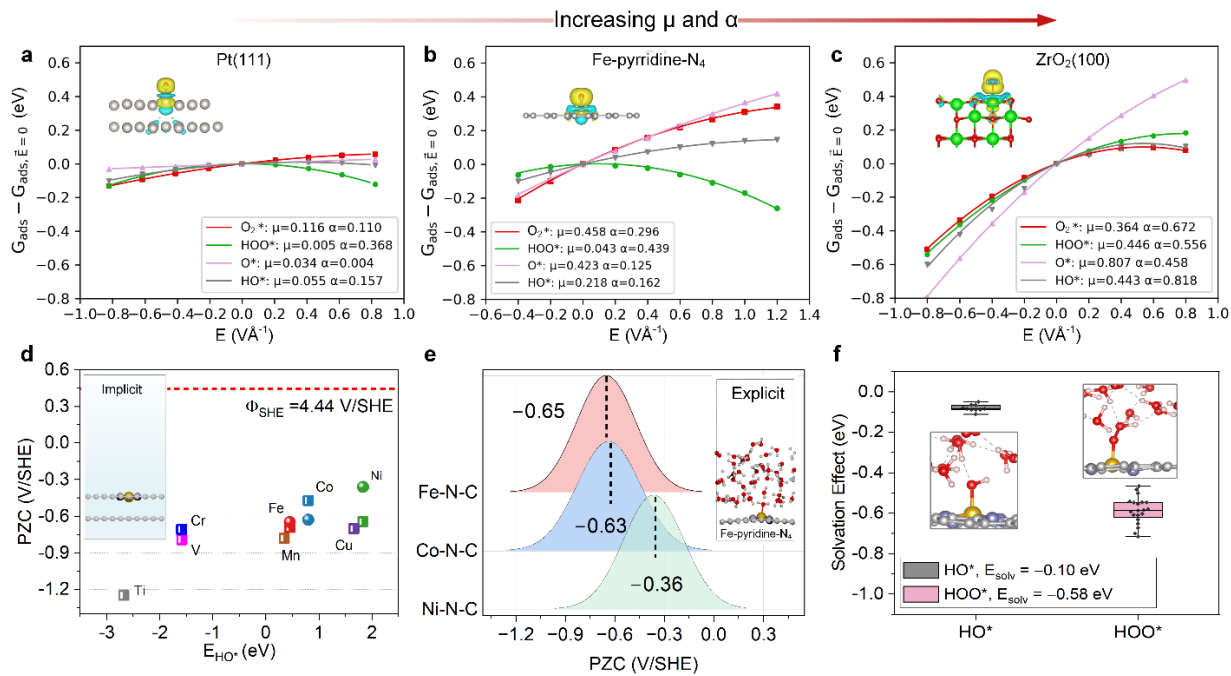

**Figure 3. Field effects and other critical factors that determine the ORR activities.** (**a**)-(**c**), Electric field effects on the adsorption free energies of ORR adsorbates with the fitted values for μ (dipole moment, $e$Å) and α (polarizability, $e^2$ V$^{-1}$) for (**a**) Pt(111), (**b**) Fe-pyridine-N$_4$, and (**c**) ZrO$_2$(100), respectively. Insets: the charge density difference induced by the adsorption of atomic oxygen. Yellow and teal colors in the isosurfaces represent electron charge gain and loss, respectively. Red, gray, and green spheres represent O, C, and Zr, respectively. The other critical factors include (**d**) the PZCs calculated from implicit and (**e**) explicit models and solvation effects on the (**f**) binding energy changes of HO* and HOO* in the presence of explicit water molecules. The normal distributions depicted in (**e**) originate from over 1,000 steps of the catalyst-water AIMD simulations and the box-plots in (**f**) are derived from at least 20 distinct surface-adsorbate combinations.

The PZC serves as a pivotal parameter dictating the electrochemical interactions at the metal-water interface. For example, it profoundly modulates the magnitude of the electric field for a given potential, subsequently dictating the configuration of the electric double layer.[38] In the present study, we evaluated the PZCs of selected M-N-C catalysts, comparing data derived from implicit models (**Figure 3d**) with those from AIMD simulations (**Figure 3e**). Our findings manifest a notable congruence between the PZCs of Fe-pyridine-$N_4$ and Co-pyridine-$N_4$ as determined by the VASPsol implicit model and those extrapolated from AIMD.[39] Contrarily, the PZCs of Ni-pyridine-$N_4$, when determined by both methods, exhibit a deviation of 0.3 V/SHE. Throughout AIMD simulations, the Ni-pyridine-$N_4$ surface demonstrates a minimal interaction with water molecules, whereas the Fe/Co-pyridine-$N_4$ surfaces potentially adsorb a water molecule onto the central metal atom (refer to **Supplementary Movies 1-3** for a comprehensive view). This suggests that the implicit model might possess inherent limitations in accurately describing PZCs when water molecules establish distinct interactions with material surfaces. According to our computational results, the PZCs of M-N-C catalysts predominantly lie within a range of −0.3 to −0.6 V/SHE. This range also contrasts with the PZCs of TM surfaces, which are conventionally observed to be positive.[35] In pursuit of a unified model for this research, we employed an average value of the PZCs deduced from the AIMD simulations to derive microkinetic activity models.

Solvation effects are also important for electrochemical reactions that take place at the electrode−electrolyte interface.[40] In this work, we analyze the influence of explicit solvation on the stability of adsorbed intermediates during ORR on M-N-C catalysts. Previous research has reported explicit solvent stabilization of HO* on Pt(111), with reported values of 0.1~0.3 eV depending on coverage.[26] Tripkovic *et al.*[41] found HOO* to be stabilized by 0.5 eV using a half-dissociated water layer network on Pt(111). Liu *et al.*[42] found that O* and $O_2$* experience minimal stabilizations, while HO* and HOO* are stabilized by approximately 0.6 and 0.7 eV, respectively. Notably, although the solvation effects of these oxygen species have been extensively examined on PGMs, the extent to which adsorbates on M-N-C catalysts are affected by solvation remains a less explored area. Herein, we examine the solvation effects on the Fe-pyridine-N catalyst surface through AIMD simulations using explicit water molecules. For each adsorbate, we uniformly sampled 20 substrate-adsorbate combinations from a pool of >1,000 steps generated by AIMD simulations (other details are shown in **Supplementary Methods**). As the results, we apply a mean value of −0.1 and −0.58 eV to adjust the adsorption energies of HO* and HOO*, respectively, to model solvation effects.

Based on our comprehensive analysis that combined all pertinent factors mentioned above, we have formulated pH-dependent microkinetic models for M-N-C catalysts. Distinct from the ORR volcano plots previously observed for TMs or TMXs,[25, 35, 36] the ORR volcano of M-N-C SACs exhibits a unique dynamic evolution from a single peak (*i.e.*, the *Sabatier* optimum) in alkaline conditions to a double peak in acidic environments. In **Figure 4a**, the two peaks represent the $4e^-/2e^-$ activity peaks at 0.8 V/RHE, respectively. An "acid trap" creates this division between them, as seen in the same figure. For further clarification, the $2e^-$ volcanos at 0.6 V/RHE are also presented in **Figure 4b**. On the pH-dependent volcano plots, several representative M-N-C catalysts are indicated, and their pH-dependent behaviors remarkably align with previous experimental observations summarized in **Figure 1,** except for Cu/Ni-based catalysts in alkaline media, likely due to the existence of active sites other than the metal atop-site for these weaker bonding catalysts. For instance, the ORR activity of pyrrole-N-type Fe-N-C has been experimentally confirmed to be superior in both acidic and alkaline solutions.[1] Our model also reflects this, with the calculated Fe-pyrrole-$N_4$ data point (represented as squares in **Figure 4a**) meeting both the acidic and alkaline peaks concurrently. **Figure S5h** details the ORR activity of pyridine-N-type Fe-N-C, revealing that Fe-pyridine-$N_4$ exhibits reduced activity in acidic conditions compared to Fe-pyrrole-$N_4$. Furthermore, the higher $2e^-$ selectivity of Co-centered M-N-C catalysts, previously substantiated in experiments and shown in **Figure 1**, is also mirrored in our model, as shown in **Figure 4b** (upper triangles). These findings offer preliminary evidence of the model's reliability and its potential in predicting the performance of newly designed M-N-C catalysts.

For a deeper understanding, **Figure 4c-e** depicts the reaction rate analysis of elementary ORR steps on M-N-C catalysts for the $4e^-$-alkaline, $4e^-$-acidic, and $2e^-$ processes. The transition from a single peak to a double peak in these volcanos arises from the distinct rate-determining steps (RDS). When the adsorption strength of hydroxyl is strong, HO* removal is typically regarded as the RDS, consistent with the volcanoes observed on TMs and TMXs.[25] As $G_{HO*}$ shifts to more positive values, M-N-C catalysts exhibit a unique scaling compared to TMXs or TMs. Specifically, for M-N-C, the scaling relation is $E_{O*} = 1.5E_{HO*} + 1.26$. In contrast, TMXs and TMs follow $E_{O*} = 2.0E_{HO*} + 0.87$[25] and $E_{O*} = 2.0E_{HO*} + 0.20$,[36] respectively. This suggests that M-N-C catalysts have stronger O* adsorption than TMXs, yet weaker than TMs when the descriptor $E_{HO*}$ lies between 0.8~2.0 eV. Consequently, M-N-C catalysts are generally more efficient at facilitating O-O bond breaking compared to TMXs but may produce more $H_2O_2$ than TMs. Further comparisons can be found in **Figure S5**. **Figure 4c** illustrates that in alkaline solutions, the right leg of the M-N-C volcanoes is constrained by $O_2$ protonation. Conversely, in acidic solutions, the increased field response of O* and HOO* on M-N-C catalysts, combined with their lower PZCs, leads to more positive $G_{O*}$ and more negative $G_{HOO*}$. As a result, the O-O bond breaking in HOO* emerges as the primary controlling factor (**Figure 4d**). In the $2e^-$ ORR process (as depicted in **Figure 4e**), the RDS for $H_2O_2$ production is predominantly the HOO* + $H^+$ + $e^- \rightarrow H_2O_2$ step, which competes with the HOO* + $H^+$ + $e^- \rightarrow$ O* + $H_2O$ step. Essentially, the $4e^-/2e^-$ selectivity is determined by whether HOO* prefers to accept a proton-electron pair or to break the O-O bond. This observation aligns with findings from a recent AIMD study.[11] Therefore, assuming that the free energy of dissolved $H_2O_2$ remains constant with electrode potential, the pH dependence of the $4e^-/2e^-$ selectivity can be mainly attributed to the field effects on HOO* and O*. Specifically, we find that the HOO* adsorbates always display the largest polarizability among other ORR adsorbates, resulting in a significant stabilization effect in acidic solutions. This is consistent with experimental results that M-N-C catalysts generally exhibit a greater $2e^-$ ORR selectivity in acid compared to that in alkaline (**Figure 1**).[22] These discoveries reveal the fundamental reasons for the pH-dependent performance and selectivity of M-N-C catalysts.

With these enhanced understandings, we will demonstrate the strength of our model in elucidating the unique properties of M-N-C catalysts. **Figure 4f** illustrates the variations in the ORR turnover frequency (TOF) as a function of pH conditions (delineated by the blue and red solid arrows) and the descriptor $G_{HO*}$. These variations can be attributed to the interplay of distinct electric field responses and unique scaling relations. For instance, the dashed red arrow in **Figure 4f** indicates that pyridine-N catalysts exhibit lower ORR activities in acidic environments due to their relatively larger dipole moments (as detailed in **Table S5**). This nuanced observation is the essential reason for the superiority of Fe-pyrrole-N over Fe-pyridine-N in acidic ORR experiments.[1, 28] Furthermore, the intrinsic scaling relations of M-N-C catalysts critically influence the contour of the 2D pH-dependent volcano plot. As shown in **Figure 2e**, an approximately 0.5 eV lower intercept of scaling relations for unsaturated N-coordinated M-N-C catalysts (M-N$_{3/2/1}$) suggests that HOO* is more inclined to undergo O-O bond dissociation than to accept a proton-electron pair and then generate $H_2O_2$. This behavior results in a reduced $H_2O_2$ selectivity, as indicated by the blue dashed arrow in **Figure 4f**. Our model thus provides a clear explanation for the observed lower $H_2O_2$ FE in the M-N-C catalysts using standard pyrolysis methods (**Figure 1, upper left**).

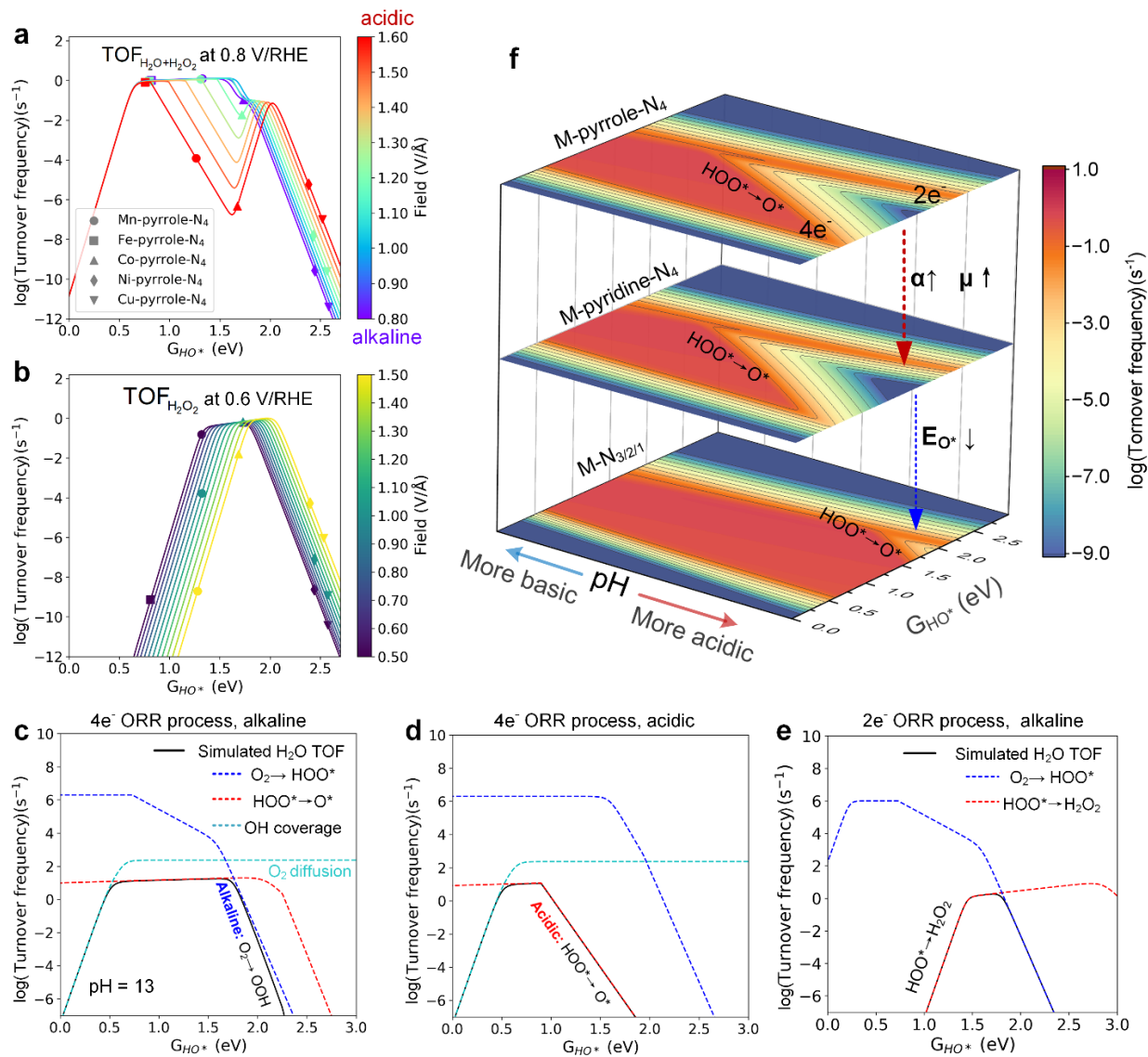

**Figure 4. Microkinetic ORR volcano models of M-N-C SACs and rate-determining analyses. a-b**, Activity volcano for the (**a**) overall ORR TOF of M–pyrrole–N catalysts at 0.8 V/RHE, (**b**) 2e$^-$ ORR TOF of M–pyrrole–N catalysts at 0.6 V/RHE. Higher electric fields correspond to lower pH values. **c-d** Rate-determining analyses of (**c**) 4e$^-$ ORR process in alkaline media, (**d**) 4e$^-$ ORR process in acids, and (**e**) 2e$^-$ ORR process in alkaline media. Dashed lines indicate the ORR activity

solved by rate-determining-step analysis with the rate limited by O₂ protonation (blue), HOO* protonation and splitting (red), and HO* coverage (cyan). (**f**) Comparative analysis of the 2D pH-dependent volcano plots derived from differing electric field responses and unique scaling relations. The M-pyridine-N$_4$ exhibits a greater sensitivity to electric fields compared to M-pyrrole-N$_3$, which results in a pronounced "acid trap" (red dashed arrow). Additionally, the M-N$_{3/2/1}$ catalysts have a distinctively lower intercept ($E_{O*}$ = 1.5$E_{HO*}$+ 0.7) than M-N$_4$ catalysts ($E_{O*}$ = 1.5$E_{HO*}$ + 1.26), contributing to a diminished 2e⁻ selectivity.

The pH-dependent model presented in this study offers a comprehensive understanding of the behaviors of M-N-C catalysts. However, when considering its potential for predicting and designing novel ORR catalysts, more rigorous validations are essential. To obtain the intrinsic activity of identically coordinated metal atoms within M-N-C catalysts, we experimentally synthesized 13 heterogeneous molecular M-N-C catalysts with precise structures. These molecular catalysts are further categorized into three groups based on central metals and adjoining functional groups: M-COF366, M-Pc, and Fe-Pc-R (for structural details, see **Figure 2** and **Figure S6-Figure S12**). We performed comprehensive characterizations on these catalysts (see details in **Supplementary Information**). The synchrotron illuminated metal K-edge X-ray absorption near-edge structure (XANES) spectra, as shown in **Figure 5a-d**, and their extended X-ray absorption fine structure (EXAFS, **Figure S12** and **Table S6**) fitting results confirmed that the metals are solely distributed in the M-N$_4$ configuration. Besides, we also assessed the ORR activity of the pristine CNT and COF366/CNT substrates (see **Figure S13** for structural details) collected in acidic and alkaline electrolytes to assess the contribution of metal-free sites, *e.g.*, doped carbon or nitrogen groups. As depicted in **Figure S14**, both substrates exhibit appreciable ORR current in alkaline electrolyte, but remain much inferior to any of the SACs assessed here, ruling out the contribution of the substrates. Besides, to offer a reasonable comparison between our experimental and theoretical results, we calculated the site-specific turnover frequency (TOF) from the $O_2$-diffusion corrected kinetic current density ($j_K$) obtained from the Koutecky-Levich equation, and they were used as the primary performance benchmarks. Additionally, we also integrated prior experimental findings from other studies.[21, 22]

Excitingly, when comparing the simulated $j_K$ in **Figure 5e** with the experimental $j_K$ in **Figure 5g**, the predicted onset potentials and curve shapes under different pH conditions are perfectly consistent with experimental results for M-COF366 series. Similar results for M-Pc series can be found in **Figure S15**. Additionally, Tafel slopes for both alkaline (solid lines) and acidic conditions (dashed lines) for the simulated and experimentally observed $j_K$ currents are illustrated in **Figure 5f** and **5h**, respectively. After further analysis of these Tafel plots, a correlation plot between the predicted and experimental Tafel slopes has been given in **Figure 5i**, demonstrating the notable consistency between the model predictions and the experimental data. Certain catalysts manifest deviations of roughly 30 mV dec$^{-1}$. These discrepancies may stem from the intrinsic pre-factors and PZCs associated with the real reactions and catalyst structures. Nevertheless, our model provides an accurate fundamental trend in the rate-limiting step alterations, attributing to the integration of transition states and field effects. Most importantly, the "acid trap" area identified by our theoretical model is further validated by experimental TOF results (detailed TOF plots can be found in **Figure S16**). As discussed previously, the ORR activities of M-N-C catalysts in an acidic environment is initially constrained by the HOO*-to-O* transition, subsequently by the HOO*-to-$H_2O_2$ transition. Interestingly, experimental data in **Figure 5j** reveals that the acidic ORR TOF of molecular M-N-C catalysts first undergoes a noticeable decline, followed by an increase attributed to $H_2O_2$ production. A comparison between simulated and experimental 2e-TOF for the M-N-C catalysts can be found in **Figure S17**. Therefore, during the design of ORR catalysts for fuel cell or $H_2O_2$ production, our model advises caution against venturing into the pH-dependent acidic trap region, where the ORR activities for both $2e^-$ and $4e^-$ are low.

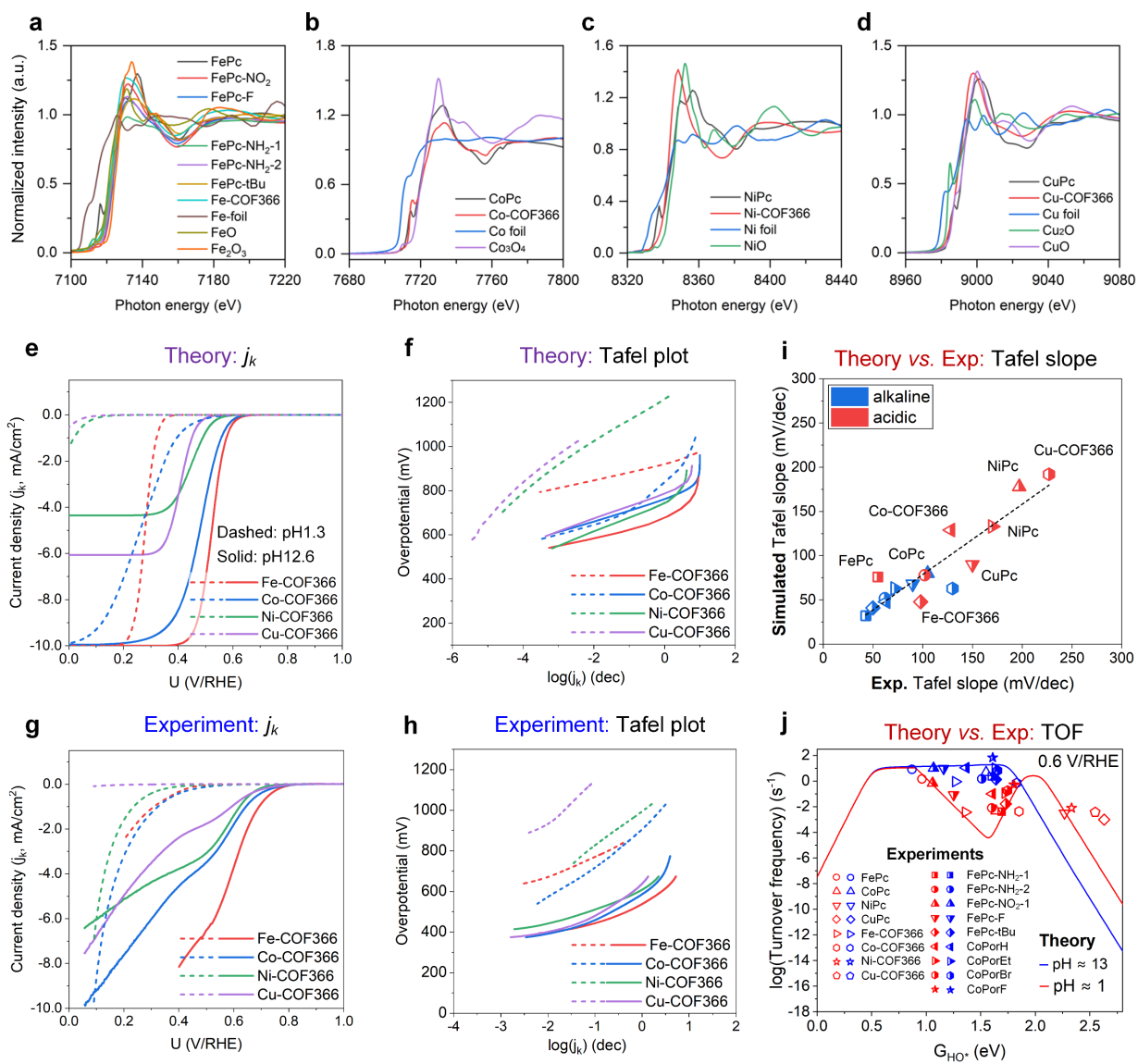

**Figure 5. Experimental characterization, performance tests, and theoretical predictions. a-e**, Metal K-edge X-ray absorption near-edge structure (XANES) of the MPc/CNT, FePc/CNT-R, and M-COF366/CNT catalysts and their reference samples. (**a**) Fe-based, (**b**) Co-based, (**c**) Ni-based, and (**d**) Cu-based. Corresponding fitted $k^3$-weighted FT-EXAFS spectra are available in **Figure**

**S12**. (**e**) Simulated $j_k$ at pH 1.3 (dashed line) and 12.6 (solid line) and (**f**) simulated Tafel plot for the typical M-COF366 catalysts. (**g**) Experimental $j_k$ at pH 1.3 (dashed line) and 12.6 (solid line) and (**h**) experimental Tafel plots for the typical M-COF366 catalysts. (**i**) Correlation plot of the experimental and simulated Tafel plots. (**j**) Turnover frequency of the molecular M-N-C catalysts obtained from experimental measurements and theoretical predictions.

## 4. CONCLUSIONS

This research offers deep insights into the pH-dependent ORR behavior of M–N–C catalysts and introduces a comprehensive model that can be potentially applied to other electrocatalytic processes. Returning to the initial key questions raised by **Figure 1**, this study presents a unified model that comprehensively addresses each of them and reveals that the pH dependence of M-N-C catalysts essentially originates from their moderate dipole moments and polarizability for the adsorbates O* and HOO*, as well as their unique scaling relations among ORR adsorbates. By considering all relevant factors related to the 2e⁻/4e⁻ ORR pathways, such as thermodynamics, kinetics, PZCs, and environmental corrections, we carefully identify the different rate-determining steps in acidic and alkaline media. Based on these new insights, the key to improving the acidic activity of M-N-C catalysts is to lower the energetics of O-O bond breaking in the HOO* → O* step. Promising approaches might include exploring M-N-C catalyst structures or active sites that exhibit optimal HO* adsorption energy and reduced O* dipole moments, or utilizing an external electromagnetic field to stabilize O* adsorption.

To tune the selectivity of M-N-C catalysts, our analysis suggests that the primary focus should be the comparison of the energetics of the O-O bond breaking in HOO* and the protonation barrier from *OOH to $H_2O_2$. This fundamental principle could enable rapid evaluation of the 2e⁻/4e⁻ preference of a M-N-C catalyst across various pH conditions. For more precise estimations, it is essential to consider the potential-dependent free energies and accurate PZCs. The classic *Sabatier* principle tells us that for good catalysis, the binding strengths of adsorbates shouldn't be too tight nor too loose. Adding to this, our study warns against designing M-N-C catalysts where the adsorbates are in the so-called "acid trap". In this zone, both $2e^-$ and $4e^-$ ORR are less efficient. With these findings, scientists have a clearer roadmap to tailor the behavior of M-N-C catalysts. This knowledge can fine-tune their activity based on pH, paving the way for breakthroughs in ORR catalysis and any other electrochemical reactions influenced by pH.

ASSOCIATED CONTENT

The data that support the findings of this study are included in the published article and its **Supplementary Information.** All other data are available from the authors upon reasonable request. In addition to being available upon request, an online database on M-N-C catalysts in ORR can be accessed via this URL: https://m-n-corrdatabase-ex9atnoev1e.streamlit.app/.

The following files are available free of charge.

> pH-dependent summary of M-N-C catalysts, detailed computational and experimental computational methods (PDF), Intrinsic dipole moment (μ) and polarizability (α) of M-pyridine/pyrrole-N catalysts, pH-Dependent volcano models for typical catalysts, experimental benchmark for the 2e- ORR kinetic volcano plot, kinetic current densities and Tafel slope fitting for MePc/CNT, characterization and Electrochemical Performance of the CNT and COF366/CNT substrate, Microstructural characterization results, metal loading determined by ICP-AES, EXAFS fitting results, experimental turnover frequency (PDF).
>
> Supplementary Movie 1-3 (MOV)

AUTHOR INFORMATION

**Corresponding Author**

**Li Wei** - School of Chemical and Biomolecular Engineering, The University of Sydney, Darlington, New South Wales, 2006, Australia. Email: l.wei@sydney.edu.au.

**Hao Li -** Advanced Institute for Materials Research (WPI-AIMR), Tohoku University, Sendai, Japan. E-mail: li.hao.b8@tohoku.ac.jp.

**Authors**

**Di Zhang -** Advanced Institute for Materials Research (WPI-AIMR), Tohoku University, Sendai, Japan**;** State Key Laboratory of Mechanical System and Vibration, Shanghai Jiao Tong University, Shanghai 200240, China; E-mail: zhangdi2015@sjtu.edu.cn.

**Zhuyu Wang** - School of Chemical and Biomolecular Engineering, The University of Sydney, Darlington, New South Wales, 2006, Australia; E-mail: zwan8100@uni.sydney.edu.au.

**Fangzhou Liu** - School of Chemical and Biomolecular Engineering, The University of Sydney, Darlington, New South Wales, 2006, Australia; E-mail: fliu2578@uni.sydney.edu.au.

**Peiyun Yi -** State Key Laboratory of Mechanical System and Vibration, Shanghai Jiao Tong University, Shanghai 200240, China; E-mail: yipeiyun@sjtu.edu.cn.

**Linfa Peng -** State Key Laboratory of Mechanical System and Vibration, Shanghai Jiao Tong University, Shanghai 200240, China; E-mail: penglinfa@sjtu.edu.cn.

**Yuan Chen** - School of Chemical and Biomolecular Engineering, The University of Sydney, Darlington, New South Wales, 2006, Australia; E-mail: yuan.chen@sydney.edu.au.

Notes

The authors declare no competing financial interest.

ACKNOWLEDGMENT

This research was supported by JSPS KAKENHI (No. JP23K13703) and the Iwatani Naoji Foundation. We acknowledge the Center for Computational Materials Science, Institute for Materials Research, Tohoku University for the use of MASAMUNE-IMR (No. 202212-SCKXX-0204) and the Institute for Solid State Physics (ISSP) at the University of Tokyo for the use of their supercomputers. Hao Li and Li Wei acknowledge the financial and technical support provided by the University of Sydney under the International SDG Collaboration Program, the Australian Centre for Microscopy & Microanalysis (ACMM), and the Sydney Informatics Hub (SIH), and also acknowledge the computational resources provided by the National Computational Infrastructure (NCI). Li Wei acknowledges the funding support provided by the Australian Research Council Future Fellowship (Grant No. ARC-FT210100218). Di Zhang acknowledges the support of the China National Postdoctoral Program for Innovative Talents from the China Postdoctoral Science Foundation (No. BX2021178) and National Natural Science Foundation of China (No. 22309109). Di Zhang gratefully acknowledges the support provided by the Shanghai Jiao Tong University Outstanding Doctoral Student Development Fund and Siyuan-1 cluster supported by the Center for High Performance Computing at Shanghai Jiao Tong University.

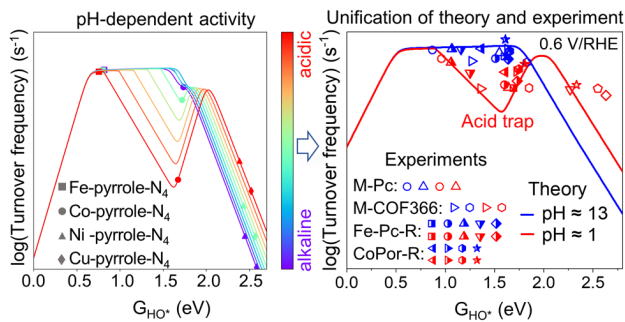

TOC Graphic